\documentclass[preprint,showpacs,amssymb,aps,nofootinbib]{revtex4}

\usepackage{amssymb,amsmath}
\usepackage{amssymb}

\usepackage{slashed,bbold}
\newcommand{\citen}{\cite}

\begin{document}

\title{\small EQUIVALENCE OF THE PATH INTEGRAL FOR FERMIONS
 \\
IN CARTESIAN AND SPHERICAL COORDINATES}

\author{
 Horacio E. Camblong,$^{1}$ 
and 
Carlos R. Ord\'{o}\~{n}ez$^{3}$}

\affiliation{
$^{1}$ Department of Physics, 
University of San Francisco, San Francisco, California 94117-1080 \\
$^{2}$   Department of Physics, University of Houston, Houston,
TX 77204-5506}

\begin{abstract}
The path-integral calculation for the free energy 
of a spin-1/2 Dirac-fermion gas
 is performed in spherical polar coordinates
for a flat spacetime geometry.
 Its equivalence with the
 Cartesian-coordinate representation is explicitly established.
This evaluation involves a relevant limiting  
case of the fermionic path integral in a Schwarzschild background, whose near-horizon limit 
has been shown to be related to black hole thermodynamics. 

\keywords{Path integral quantization; Dirac fermion field; finite-temperature field theory; coordinate invariance.}
\end{abstract}

\pacs{11.10.-z, 11.10.Wx}

\maketitle

\section{Introduction}

  In this paper we develop an
  expansion in fermionic spherical harmonics 
  in flat spacetime that serves as a nontrivial test for the technical correctness 
  of a subtle fermionic calculation of black hole thermodynamics.
  The original context is provided by the black-hole entropy calculation
 in the framework of `t Hooft's brick-wall approach~\cite{BW_t-Hooft}, which
 can be extracted in a systematic manner 
  via a near-horizon (NH) expansion technique.
  This methodology
not only shows that black-hole thermodynamics is associated with 
the leading  divergent order of the entropy calculation
but it also  
provides a clear origin for 
the Bekenstein-Hawking area law~\cite{BH_thermo_CQM,semiclassical_BH_thermo,PI-fermion_curvedST}
from 
a NH conformal symmetry.
In Ref.~\citen{BH_thermo_CQM},
an expansion in spherical harmonics 
was used 
for the reduction to a radial one-dimensional semiclassical analysis of the density of modes
 for spin-zero fields.  
Even though  this does not exhaust the arsenal of available techniques---NH methods without 
the need for an expansion in spherical harmonics
are also known for the scalar case~\cite{semiclassical_BH_thermo,
holographic_scaling_I}---the NH origin of the thermodynamics
is most obvious with the resolution in spherical coordinates.
The appropriate generalization of Ref.~\citen{BH_thermo_CQM}
 in fermionic spherical harmonics 
can be derived~\cite{PI-fermion_curvedST}
in both the canonical  and path-integral frameworks,
 using the same techniques presented below for the flat spacetime case.
In addition to the subtleties discussed in the following sections,
in the case of a
black hole background,
the NH method in spherical coordinates  
proves to be a crucial ingredient in the 
 extraction of the Bekenstein-Hawking area law.

 In essence, 
  while the path-integral calculation of the free energy for a free-fermion gas
 in flat spacetime is well known using Cartesian coordinates~\cite{TFT1a,TFT1b,TFT2},
its counterpart in spherical coordinates is not. 
Thus, through a series of steps leading to the free-energy path integral
in spherical coordinates,
 we complete this calculation 
and explicitly show the equivalence 
with Cartesian coordinates---also providing
  a strong check on the correctness 
 of the corresponding fermionic path-integral  in more general spherically symmetric backgrounds,
 including the background leading to black hole thermodynamics.
 Moreover, this calculation highlights some conceptual and technical 
 properties that are crucial for similar computations in curved spacetime, and lends itself to generalizations 
 for higher spin fields.

In its greatest generality,
the analysis of Dirac operators in spherical coordinates can be fully implemented within
the Newman-Penrose geometric
framework~\cite{NP-formalism}
and the use of spin-weighted spherical 
harmonics~\cite{spin-weightedY_Newman-Penrose1, spin-weightedY_Newman-Penrose2, spin-weightedY_Newman-Penrose3}.
These techniques have led to solutions of remarkable generality with Chandrasekhar's 
work~\cite{Chandrasekhar}.
There is considerable interest, however, in 
the development of additional techniques for specific problems;
for example,
 for problems in the background of gravitational fields~\cite{Shishkin_Dirac-gravity} as well as
in curvilinear orthogonal coordinates in flat spacetime, mainly by 
 algebraic methods of separation of 
 variables~\cite{Shishkin-Villalba_separation1,Shishkin-Villalba_separation2, Shishkin-Villalba_separation3},
to name just a few.
Our interest is in
 the use and development of techniques 
that simplify the path integral treatment for specific problems with Dirac fermions.
In this paper, we focus on the case of spherical polar coordinates, and adapt
some of these techniques 
to our path integral problem.
Thus,
 we follow a 
remarkably simple constructive approach specifically tailored for the computation of determinants and 
statistical properties of the Dirac operator.
This approach leads to the generalized spherical harmonics discussed in Sec.~\ref{sec:PI_spherical-coordinates}
and in App.~\ref{sec-app:spherical_reduction}.

 With
 these considerations in mind, we first
 turn 
 to  the setup of the problem 
 in both coordinate systems, and then to the 
development of the  relevant fermionic spherical harmonic expansion leading to their equivalence.

\section{The Fermion Problem in Flat Spacetime: 
\\
Euclideanized Action and Partition Function}

Consider the Euclideanized action for free fermion fields in flat spacetime~\cite{TFT1a,TFT1b,TFT2},
\begin{equation}
S_{E} =
\int _{0}^{\beta }
d\tau  
\int d^{3} x 
\,
\bar{\psi }
\,
\left( \tau , \mathbf{x} \right)
\,
\left( \gamma_{E}^{\mu } 
\partial _{\mu }
 + m\right)
 \,
 \psi \left(\tau , \mathbf{x} \right)
 \, ,
 \label{eq:Euclidean-action}
\end{equation} 
 where the  Euclidean time is  $\tau = i t$ and the 
 relevant conventions and definitions, including those of the Euclidean Dirac matrices 
 $\gamma_{E}^{\mu } $ 
   are summarized in App.~\ref{sec-app:Euclidean_Dirac}.
 It should be noticed that the derivatives 
 $\partial _{\mu }$ in Eq.~(\ref{eq:Euclidean-action}), by abuse of notation,
 already involve the Euclidean time $\tau $ (this will have no further consequences 
 for the remainder of the paper, but the subtleties are absorbed in the conversion to
 $\gamma_{E}^{\mu } $).
Correspondingly, the partition function for this system is
\begin{equation}
Z=
\int_{\scriptsize
\begin{array}{l} 
{\psi \left(0,  \mathbf{x} \right)
=
-\psi \left(\beta ,\mathbf{x} \right)}
 \\
 {\bar{\psi }
\left(0, \mathbf{x} \right)
=-\bar{\psi }\left(\beta , \mathbf{x} \right)} 
 \end{array} 
 }
 {\mathcal D}
\bar{\psi }
\left(\tau , \mathbf{x} \right)
 {\mathcal D}
 \psi \left(\tau ,\mathbf{x} \right) 
\exp \left\{-S_{E} 
\right\}
 \label{eq:Z_TFT} 
\, . 
\end{equation} 
In effect~\cite{TFT1a, TFT1b,TFT2},
 for the transition to the thermal field theory, the evolution of the system in
imaginary time  $\tau$ is subject to appropriate 
boundary conditions: 
for fermionic fields, anti-periodic boundary conditions should be satisfied
with respect to $\tau$, i.e.,
\begin{equation}
\left\{
\begin{array}{l} 
{\psi \left(0, \mathbf{x} \right)
=
-\psi \left(\beta ,\mathbf{x} \right)}
 \\
 {\bar{\psi }
\left(0, \mathbf{x} \right)
=-\bar{\psi }\left(\beta ,\mathbf{x} \right)} 
\, ,
 \end{array} 
\right.
\end{equation} 
where $\beta =1/T$ is the inverse temperature.
This implies that the frequencies associated with the Euclidean time $\tau $,
known as fermionic Matsubara frequencies, are discrete and of the form
\begin{equation}
\omega _{n} =\frac{2\pi }{\beta } \left(n+\frac{1}{2} \right)
\, ,
\;
n \in {\mathbb Z}
\, .
 \label{eq:Matsubara-frequencies}  
\end{equation} 

The translational invariance of flat spacetime
permits the  introduction
of  the momentum-space Fourier representation for the spatial part of the fermion field
\begin{equation}
\psi \left(\tau , \mathbf{x} \right)
=
\displaystyle\sum _{n}
\frac{
e^{-i\omega _{n} \tau} 
}{\sqrt{\beta } } 
 \int \frac{d^{3} p }{
\left[ \left(2\pi \right)^{3/2} 
 \sqrt{2
 \omega _{\mathbf{p} } } \right]
  }
 \,
 e^{i  \mathbf{p}\cdot \mathbf{x} }
\,
  \psi _{n} \left(\mathbf{p}\right)
  \, ,
   \label{eq:Qfield_Fourier} 
\end{equation} 
where
$\omega _{\mathbf{p} }
 = \sqrt{ \mathbf{p}^{\, 2} + m^{2} }$,
 the sum is $\sum_{n} = \sum_{n=-\infty}^{\infty}$,
 and the normalization
  corresponds to the one-particle wave function 
  $\left\langle 0 | 
  \psi \left(\tau ,\mathbf{x}\right)
   | \omega_{n}, \mathbf{p}
    \right\rangle 
   =
   e^{i(\mathbf{p}\cdot \mathbf{x} - \omega_{n}\tau)}/\sqrt{ (2\pi)^{3} 2 \omega_{\mathbf{p} } }$.
   Notice that the basic modes
   involve the functions $e^{-i P \cdot X}$.
Then, the  Euclideanized action  becomes
\begin{equation}
S_{E} =
\displaystyle\sum _{n}
\int
 \frac{d^{3} p }{  2\omega _{\mathbf{p} }  } 
 \;
\bar{\psi }_{n} \left(\mathbf{p}\right)
\,
\left(- i\gamma_{E}^{\mu } P_{\mu } 
+m\right)
\,
  \psi _{n} 
\left(\mathbf{p}\right), 
 \label{eq:Euclidean-action_p-space}
\; .
 \end{equation}  
 This defines the Euclidean Dirac operator $\mathfrak{D}_{P}=\left(- i\gamma_{E}^{\mu } P_{\mu } 
+m\right)$ in the momentum representation, with the conventions of App.~\ref{sec-app:Euclidean_Dirac}.
Therefore, using the well-known result for
Berezin path integrals with 
  Grassmann (anti-commuting)
  variables~\cite{TFT1a, TFT1b,TFT2,Grassmann},
\begin{equation}
\int \left\{\prod _{i}da_{i}^{*} da_{i} 
 \right\} 
\exp \left(-a_{i}^{*} M_{ij} a_{j} \right) 
{=} 
\det \left(M\right)
\, , 
 \label{fermionic-det_Grassmann} 
 \end{equation} 
the partition function~(\ref{eq:Z_TFT})
yields
\begin{equation}
\begin{array}{rcl} 
{Z} & {=} 
& C
\displaystyle\displaystyle\prod_{n,\mathbf{p}}
\det 
\left(- i\gamma _{E}^{\mu } P_{\mu } +m
\right) 
\\ {} 
& {=} & 
C
\left[ 
\displaystyle\prod_{n,\mathbf{p}}
\det 
\left(
- i\gamma_{E}^{\mu } P_{\mu }
 +m
\right)
 \det 
 \left(
 i\gamma_{E}^{\mu } P_{\mu } 
 +
 m
 \right)
 \right]^{1/2} 
\, , 
 \end{array} 
\end{equation} 
with $C$ a constant.  
Here we used 
\begin{equation} 
\left|\det D\right|
=
\sqrt{\det D\det D^{\dag } }
\, 
\label{eq:square-root_operator_det-relation}
\end{equation} 
for each operator in the factorization above.
Now,
in the Euclidean slash notation, 
$
\slashed{P}_{E}=
\gamma_{E}^{\mu } P_{\mu }
$,
and using the Euclidean Clifford algebra relations~(\ref{eq:Euclidean_Clifford-algebra}),
we have 
$
(i \slashed{P}_{E} + m )
(- i \slashed{P}_{E} + m)
=
\slashed{P}_{E} \slashed{P}_{E}
 + m^{2}
 =
\left(  P^{2} + m^{2} \right)
 \,  {\mathbb {1}}_{4} $,
 where 
 $  {\mathbb {1}}_{4} $
 is the 
$4 \times 4$ identity matrix.
Thus, up to a phase, we get  
\begin{equation}
{Z} 
=
\displaystyle\prod _{n,\mathbf{p}}
\det\,^{1/2} 
\left[
\left(\omega _{n}^{2} +\mathbf{p}^{\; 2}
 +m^{2}  \right) 
   {\mathbb {1}}_{4} 
  \right]
 \, ,
\end{equation}
as 
$P^2 \equiv \delta^{\mu \nu}  P_{\mu } P_{\nu} 
= 
\omega _{n}^{2} +\mathbf{p}^{\; 2}$.
Notice that 
 the calculation of the determinant involves 
products with respect to
the Dirac matrices in addition to its functional
nature.
As a result of  the simple factorization above as a direct product,
the matrix part  
leads to an overall exponent of
 4 for the functional determinant,
 so that
\begin{equation} 
\begin{array}{rcl} {Z} 
& {=} & 
\displaystyle\prod _{n,\mathbf{p}}
\left[
\det\, \! ^{1/2} 
\left(\omega _{n}^{2} +\mathbf{p}^{\; 2}
 +m^{2} 
 \right)
 \right] ^{4} 
 \\ 
 {} & {=} &
  {\displaystyle\prod _{n,\mathbf{p}}
  \det\, \! ^{2} \left(\omega _{n}^{2} 
 +\mathbf{p}^{\; 2} +m^{2} \right) .}
  \end{array} 
  \label{eq:Z_momentum}
\end{equation}  
This derivation highlights the fact that $Z$ in Eq.~(\ref{eq:Z_momentum})
is the momentum factorization (up to a constant) of
 the square of the
 $\det \left(  \square _{E} + m^{2} \right)$,
 where 
$\square _{E} =- \left( \partial _{\tau }^{2} + \boldsymbol{\nabla }^{2} \right)$
(Euclidean Klein-Gordon operator, but with the eigenvalues to be evaluated with the fermionic 
Matsubara frequencies).

 Finally, from Eq.~\eqref{eq:Z_momentum},  
 it is straightforward to get the free energy using  $F= -\ln Z/\beta$,
 as discussed in the standard references~\cite{TFT1a,TFT1b,TFT2}.

\section{Path Integral in Spherical Coordinates:
\\
 Explicit Calculation and 
Coordinate Invariance}
\label{sec:PI_spherical-coordinates}

Our construction begins with 
the familiar 
transformation
from Cartesian to spherical coordinates,
for which  the Dirac operator in the action
turns into
\begin{equation}
\mathfrak{D} 
=
\gamma_{E}^{0} \partial _{0} 
+\slashed{e}_{r} \partial _{r} 
+\slashed{e}_{\theta } \frac{1}{r} \partial _{\theta } 
+\slashed{e}_{\phi } \frac{1}{r\sin \theta } \partial _{\phi } 
+m
\, , 
\label{eq:Dirac-operator_spherical}
\end{equation} 
where, in the Euclidean slash notation, 
\begin{equation}
\slashed{e}_{i}
 =\gamma_{E}^{\mu } e_{\mu i} 
,\,\,i=r,\theta ,\phi 
\, ,
\label{eq:slashed-vierbein}
\end{equation} 
with
$
e_{\mu i}
$ being the spatial
transformation matrix with columns $\hat{e}_{i}$ 
($i=r,\theta ,\phi$)
for $\mu =1,2,3$ representing the Cartesian axes; in essence,
$\slashed{e}_{i}$ is the corresponding ``spherical component'' (i.e., rotated)  of the Gamma matrices.
For our derivation of determinants below,
it should be noticed that the last term 
is explicitly 
proportional to the $4 \times 4$ identity matrix, i.e., of the form 
$m \, {\mathbb {1}}_{4} $.

This selection of the Cartesian frame,
i.e., 
 the fixed ``Cartesian gauge,'' does not 
require the explicit use of the spin connection; thus, the derivatives
$\partial_{\mu}$ above are just 
ordinary derivatives. 
The alternative 
approach with covariant derivatives is outlined at the end of App.~\ref{sec-app:Euclidean_Dirac}.

The operator $\mathfrak{D}$, 
defined above in the original field representation
 $\psi \left(\tau , \mathbf{x} \right)$,
  is not convenient for our calculational purposes.  
 Instead,
 a more ``friendly'' geometrical version results if we perform a unitary transformation on $\psi$ 
 defined by~\cite{Shishkin-Villalba_similarity1, Shishkin-Villalba_similarity2}
\begin{equation} 
\psi \left(\tau , \mathbf{x} \right)
=
U\tilde{\psi }
\left(\tau , \mathbf{x} \right)
 \; ,
 \; \; \; \; \; \;
  \bar{\psi }\left(\tau , \mathbf{x} \right) 
 =
 \bar{\tilde{\psi }}
 \left(\tau , \mathbf{x} \right)
 U^{\dag } 
 \, ,
\end{equation} 
with the rotation 
\begin{equation}
\begin{array}{rcl} 
{U \left(
R_{z} \left(\phi \right) 
R_{y}  \left( \theta \right)
\right)
} & {=} & 
{U
\left(R_{z} 
\left(\phi \right)\right)
U \left(R_{y} \left(\theta \right)\right)
}
 \\ 
 {} & {=} &
  {
  \exp 
  \left[
  - \!
  \mbox{ \Large $\frac{i \phi }{2}$} 
 \left(\begin{array}{cc} {\sigma _{3} } & {0} 
 \\
  {0} & {\sigma _{3} }
  \end{array}
  \right)
 \right]
 \exp { 
 \left[ 
 - \!
 \mbox{ \Large $\frac{i \theta }{2}$} 
  \left(
  \begin{array}{cc}
   {\sigma _{2} } & {0} 
  \\ 
  {0} & {\sigma _{2} } 
  \end{array}
  \right)
  \right]
  \; ,
  }}
\end{array} 
 \label{eq:rotation}
\end{equation} 
where $\sigma_{j}$ ($j = 1,2,3$) are the Pauli matrices.
Geometrically,
this amounts to a realignment of the coordinate axes with
the chosen curvilinear coordinates $(r,\theta,\phi)$.
Under this field redefinition,\footnote{This transformation leaves the path integral measure invariant.}
 the Euclideanized Lagrangian becomes
\begin{equation} 
\bar{\psi }
\left(\tau , \mathbf{x} \right)
\mathfrak{D}
 \psi \left(\tau , \mathbf{x} \right)
= 
\tilde{\bar{\psi }}
\left(\tau , \mathbf{x} \right)U^{\dag }
 \mathfrak{D} U\tilde{\psi }\left(\tau , \mathbf{x} \right)\equiv \tilde{\bar{\psi }}
 \left(\tau , \mathbf{x} \right)\tilde{\mathfrak{D} }\tilde{\psi }\left(\tau , \mathbf{x} \right)
 \; , 
\label{eq:transformed_Dirac-Lagrangian} 
\end{equation} 
where, from Eq.~(\ref{eq:rotation}),
 the unitarily transformed Dirac operator,
$\tilde{\mathfrak{D} }
=
U^{\dagger}
\mathfrak{D}
U
$,
takes the form
\begin{equation} 
\tilde{\mathfrak{D} }
=\gamma_{E}^{0} \partial _{0} 
+\gamma_{E}^{3} \left(\partial _{r} +\frac{1}{r} \right)
+\gamma_{E}^{1} \frac{1}{r} \left(\partial _{\theta }
 +\frac{1}{2} \cot \theta \right)
 +\gamma_{E}^{2} \frac{1}{r\sin \theta } \partial _{\phi } 
+m \,  {\mathbb {1}}_{4}
\, ,
 \label{eq:Dirac-operator_angular-mod} 
\end{equation} 
and we used  the fact that spatial rotations
[in our case, $U$ in Eq.~(\ref{eq:rotation})]
commute with $\gamma_{E}^{0} $.
Effectively, this choice leads to the selection of self-adjoint operators associated with the 
given coordinates,
including the {\em extra terms\/}, i.e.,
${\displaystyle
\frac{1}{r}
 \; {\rm and} 
 \;
 \frac{1}{2} \cot \theta }$
for the generalized radial and polar momenta.~\footnote{This is a
modified version of the   ``rotating diagonal gauge'' 
of Refs.~\cite{Shishkin-Villalba_similarity1, Shishkin-Villalba_similarity2}.}
It should be noticed that 
these are precisely the terms generated by the spin connection in the tetrad formalism, as sketched 
in App.~\ref{sec-app:Euclidean_Dirac}.

With our representation of the Dirac matrices
(see App.~\ref{sec-app:Euclidean_Dirac}),
we explicitly have
\begin{equation} 
\begin{array}{l}
\tilde{\mathfrak{D} }
= 
{\left(
\begin{array}{cc}
 {{\mathbb {1}}_{2} }  & {0} 
\\ {0} & {-{\mathbb {1}}_{2} }
 \end{array}
 \right)
 \partial _{0} 
 -
 i
 \left(\begin{array}{cc} 
 {0} & {\sigma _{3} } 
 \\ 
 {-\sigma _{3} } & {0} 
 \end{array}\right)
 \left(\partial _{r} 
 +
 \mbox{\large $\frac{1}{r}$} 
 \right)}
  \\ 
-
   \mbox{\Large $\frac{i}{r}$ } \!
  \left(\begin{array}{cc} 
  {0} & {\sigma _{1} } 
  \\ 
  {-\sigma _{1} } & {0} 
  \end{array}
  \right)
  \left(
  \partial _{\theta }
   +\frac{1}{2} \cot \theta 
   \right)
   -
   \mbox{\Large $\frac{i}{r}$ } \!
   \left(\begin{array}{cc}
    {0} & {\sigma _{2} }
     \\
      {-\sigma _{2} } & {0} 
     \end{array}\right)
     \! \mbox{\Large $\frac{1}{\sin \theta } $}
      \partial _{\phi }
      +
      m \, {\mathbb {1}}_{4}
      \, . 
      \end{array} 
\label{eq:Dirac-operator_angular-explicit} 
\end{equation} 
 It is now necessary to attempt a separation of variables to reduce the problem to one (radial)
 dimension.  
 In particular, owing to the symmetry of the problem in angular variables, suggest an expansion 
 of the form\footnote{For notational simplicity, we are omitting the $j$ and $m_j$
  labels on the spherical functions $Y_{\left(1,2\right)}$ in 
 Eqs.~(\ref{eq:spherical_reduction})--(\ref{eq:eigenvalue-eq_sphericalH}).}
\begin{equation}
\tilde{\psi} 
\propto
\frac{e^{-i\omega _{n} \tau } }{\sqrt{\beta } }
 \left(\begin{array}{c} 
 {A_{n} (r)Y_{1} (\theta ,\phi )} 
 \\
  {B_{n} (r)Y_{2} (\theta ,\phi )}
  \\ 
  {C_{n} (r)Y_{1} (\theta ,\phi )} 
 \\ 
 {D_{n} (r)Y_{2} (\theta ,\phi )} 
 \end{array}
 \right)
 \, ,
 \label{eq:spherical_reduction}
\end{equation} 
to be used along with
Eqs.~(\ref{eq:transformed_Dirac-Lagrangian})--(\ref{eq:Dirac-operator_angular-explicit}).

Following App.~\ref{sec-app:spherical_reduction},
 we introduce the 
``generalized spherical harmonic'' eigenfunctions $Y_{1}$ and $Y_{2}$,
\begin{equation}
\left(
\left[
\begin{array}{cc} 
{0} & {1} 
\\
 {1} & {0} 
 \end{array}
 \right]
 \left(\partial _{\theta } 
 +\frac{1}{2} \cot \theta \right)
 +\left[\begin{array}{cc} 
 {0} & {-i} 
 \\ 
 {i} & {0} 
 \end{array}
 \right]
\frac{1}{\sin \theta } 
 \partial _{\phi } 
 \right)
 \left(\begin{array}{c}
  {Y_{1} } \\ {Y_{2} } 
  \end{array}
  \right)
  =
  \left(
  \begin{array}{c} 
  {\lambda _{+} Y_{1} } 
  \\ {\lambda _{-} Y_{2} }
   \end{array}
  \right) 
  \, ,
   \label{eq:eigenvalue-eq_sphericalH}
\end{equation} 
where
\begin{equation} \label{1.22)} 
\lambda _{\pm } =\pm \left(j+1/2\right). 
\end{equation} 
Then, the effect of the Euclidean Dirac operator on the fields \eqref{eq:spherical_reduction} is 
\begin{equation} \label{1.21)} 
\tilde{\mathfrak{D} } \tilde{\psi}
 =
 \left(
 \begin{array}{c}
  {-i
  \left[
  \left(\partial _{r} +\frac{1}{r} \right)C+\frac{\lambda _{+} }{r} D
  \right]
  Y_{1}
   +
   \left(-i\omega _{n} +m\right)AY_{1} } 
   \\ 
   {-
   i
   \left[
   -\left(\partial _{r} 
   +\frac{1}{r} \right)D+\frac{\lambda _{-} }{r} C
   \right]
   Y_{2} 
   +
   \left[
   -i\omega _{n} 
   +m
   \right]
   BY_{2} }
    \\ 
    {i
    \left[
    \left(\partial _{r} +\frac{1}{r} \right)A
   +\frac{\lambda _{+} }{r} B
   \right]
   Y_{1} 
   +\left(i\omega _{n} 
   +m\right)CY_{1} } 
   \\ {i\left(-\left(\partial _{r} 
   +\frac{1}{r} \right)B+\frac{\lambda _{-} }{r} A\right)Y_{2} 
   +\left(i\omega _{n} 
   +m\right)
   BY_{2} } 
   \end{array}\right)
   \frac{e^{-i\omega _{n} \tau } }{\sqrt{\beta } } 
   \, . 
\end{equation} 
We use the above harmonics to expand  the spatial part of the fermion field, 
leaving the Euclidean time as before,
\begin{equation} 
\begin{array}{l} 
{ \tilde{ \psi }
=
\displaystyle\sum _{j,m_{j} ,n} 
\frac{e^{-i\omega _{n} \tau} }{\sqrt{\beta } } 
\left(\begin{array}{c} 
{A_{nj} (r)Y_{1jm_{j} } } 
\\ 
{B_{nj} (r)Y_{2jm_{j} } } 
\\
 {C_{nj} (r)Y_{1jm_{j} } } 
\\
 {D_{nj} (r)Y_{2jm_{j} } } 
\end{array}
\right)
}
\\ 
{
\tilde{\bar{\psi }}
=
\displaystyle\sum _{j,m_{j} ,n} 
\frac{e^{ i\omega _{n} \tau}}{\sqrt{\beta } }
\Biggl(
\begin{array}{cccc} 
{A_{nj}^{*} (r)Y_{1jm_{j} }^{*} } 
  \;   \; 
  & 
{B_{nj}^{*} (r)Y_{2jm_{j} }^{*} } 
  \;    \;
  & 
{-C_{nj}^{*} (r)Y_{1jm_{j} }^{*} } 
  \;   \;
  & 
{-D_{nj}^{*} (r)Y_{2jm_{j} }^{*} }
 \end{array}
 \Biggr)
 \, .
 } 
 \end{array} 
 \label{eq:psi_bar-psi_expansions} 
\end{equation} 
Using 
\begin{equation} 
\label{1.24)} 
\begin{array}{l}
 {\iint  Y_{1jm_{j} }^{*} Y_{1j'm'_{j} } 
 \sin \theta d\theta d\phi =\delta _{jj'} \delta _{m_{j} m'_{j} } 
 }
 \\ 
 {
 \iint  Y_{2jm_{j} }^{*} Y_{2j'm'_{j} }
  \sin \theta d\theta d\phi 
 =
 \delta _{jj'} \delta _{m_{j} m'_{j} } 
 \; ,
 }
  \end{array} 
\end{equation} 
and
\begin{equation}
\frac{1}{\beta} 
 \int_{0}^{\beta} d \tau
 \, e^{-i(\omega_{n} - \omega_{n'}) \tau } = \delta_{nn'}
\, , 
\end{equation}
the Euclideanized action becomes
\begin{eqnarray}
 \label{1.25)} 
S_{E}
 =
& - &
i
\displaystyle\sum _{n,j,m_{j} }
\int  dr r^{2} 
\biggl\{
\omega _{n} 
\biggr. 
\left(
A_{nj}^{*} A_{nj} 
+B_{nj}^{*} B_{nj} 
+C_{nj}^{*} C_{nj} 
+D_{nj}^{*} D_{nj} 
\right)
\nonumber
\\
 & + & 
 \left[
 A_{nj}^{*} \left(\partial _{r}
 +\frac{1}{r} \right)C_{nj} -B_{nj}^{*} \left(\partial _{r}
  +\frac{1}{r} \right)D_{nj} 
 \right. 
 \nonumber
\\
& + &
\left. 
C_{nj}^{*} 
\left(\partial _{r} +\frac{1}{r} \right)A_{nj} -D_{nj}^{*} 
\left(\partial _{r} 
+\frac{1}{r} 
\right)
B_{nj} 
\right]
\nonumber
\\ 
& + &
 \mbox{\Large $\frac{ \left(j+ 1 / 2 \right) }{r }$} 
 \left(A_{nj}^{*} D_{nj} -B_{nj}^{*} C_{nj} +C_{nj}^{*} B_{nj}
-
D_{nj}^{*} A_{nj} \right)
\nonumber
\\
\biggl.
& + &
{
 im
\left(A_{nj}^{*} A_{nj} +B_{nj}^{*} B_{nj} -
C_{nj}^{*} C_{nj} -D_{nj}^{*} D_{nj} 
\right)}
\biggr\}
\, .
\end{eqnarray}  
This expression can be written as 
\begin{equation} 
S_{E}
=
\displaystyle\sum _{n,j,m_{j} }\int drr^{2}  
 \bar{\psi }_{njm_{j} } \Omega _{njm_{j} } \psi _{njm_{j} }  
\, ,
\label{eq:action_spherical} 
\end{equation} 
where (omitting the tilde notation below)
\begin{equation}
\begin{array}{l} 
{ 
{\psi} _{njm_{j} }
 =
 \left(\begin{array}{c}
  {A_{nj} }
  \\ 
  {B_{nj} }
   \\ 
   {C_{nj} } 
   \\ 
   {D_{nj} } 
   \end{array}\right)} 
   \\ 
   {
   {\bar{\psi }}_{njm_{j} }
    =
    \Biggl(
    \begin{array}{cccc} 
    {A_{nj}^{*} }  
     \;   \;  \;
     & 
     {B_{nj}^{*} }
      \;   \;  \;
       & 
       {-C_{nj}^{*} } 
        \;   \;  \;
        & 
        {-D_{nj}^{*} } 
    \end{array}
    \Biggr)
    }
    \\ 
   {\Omega _{njm_{j} }
     =
     -
     i
     \left[
     \gamma^{0} \omega _{n} 
     +\gamma^{3} 
     \left(\partial _{r} 
     +
      \mbox{\Large $\frac{ 1 }{r }$} 
   \right)
   +
     i\gamma^{2} 
        \mbox{ 
   \Large $
\frac{
\left(
j+
1/2
 \right)
}{r}    
     $}
    +
   im 
   \,  {\mathbb {1}}_{4}
   \right]
     \, .
     } 
     \end{array} 
      \label{eq:Z_spherical1} 
\end{equation} 
In  Eq.~(\ref{eq:Z_spherical1}),
the standard (Minkowskian) Dirac matrices $\gamma^{\mu}$ are restored for convenience, and will be used
for the remainder of this paper.
Thus, the partition functional integral~(\ref{eq:Z_TFT}), from Eqs.~(\ref{fermionic-det_Grassmann}),
(\ref{eq:action_spherical}) and 
(\ref{eq:Z_spherical1}),
becomes
(up to an irrelevant constant)
\begin{equation}
Z
=
\displaystyle\prod_{j,n} 
\det\,^{2j+1} 
\left[
-i\gamma^{0} \omega_{n} 
-i\gamma^{3} \left(\partial_{r} 
+\frac{1}{r} \right)+\gamma^{2} 
\frac{\left(j+1/2 \right)}{r} 
+m
\,  {\mathbb {1}}_{4}
\right]
\, ,
 \label{eq:Z_spherical2} 
\end{equation} 
where the multiplicity  $(2j+1)$ (arising from $|m_{j}| \leq j$)
is explicitly displayed.
To evaluate the partition function~(\ref{eq:Z_spherical2}),
we again use the properties of the Dirac matrices and the identity
\begin{equation}
\det \left(D\right)=\sqrt{\det \left(D\right)\det \left(D^{\dag } \right)}  
\end{equation} 
(up to a phase factor)
for each spherical component of the Dirac operator.
As a result,
\begin{equation}
Z=\displaystyle\prod_{j,n} 
\det\,^{ j+1/2 }
 \left[
 \left(
m^{2}
+
 \omega _{n}^{2} 
+ p_{r}^{2}
+\frac{ \left(j+1/2 \right)^{2} }{r^{2} } 
\right)
\,  {\mathbb {1}}_{4}
+ \frac{ \left(j+ 1/2 \right) }{r^{2} }
\begin{pmatrix} 
\sigma _{1} 
& 
0
\\ 
0 & \sigma _{1} 
\end{pmatrix}
\right]
\; ,
  \label{eq:Z_spherical3}  
\end{equation} 
where
\begin{equation}  
p_{r} 
 =
-i  \left(\partial_{r} 
     +
   \frac{ 1 }{r }
   \right)
   \; 
   \label{eq:radial-momentum}
      \end{equation} 
 is the self-adjoint radial momentum operator, with
\begin{equation}  
p_{r}^{2}  
 =
 -
   \left( \partial^{2}_{r} + \frac{2}{r} \partial_{r} \right)
 =
 -\frac{1}{r^{2} } \frac{d}{dr} \left(r^{2} \frac{d}{dr} \right)
   \; .
  \end{equation} 
Notice that the determinant symbol
``det'' in Eqs.~(\ref{eq:Z_spherical2})  and (\ref{eq:Z_spherical3})  
 involves
 a direct product of
 the radial functional attribute ($r$) and the Dirac matrix character. 
Now, the operator in Eq.~(\ref{eq:Z_spherical3})  
   can be rewritten as 
\begin{equation} 
\Lambda_{4}^{\left(j,r\right)} 
=\left[\begin{array}{cc} 
\Omega_{\left(r,j\right)}
 \,  {\mathbb {1}}_{2}
 + W(r) \,
 \sigma_{1} & {0} 
 \\ {0} 
 & \Omega _{\left(r,j\right)}
 \,  {\mathbb {1}}_{2}
 +
 W(r) \,
  \sigma_{1} 
 \end{array}\right]
 \; ,
 \label{eq:Omega-Lambda_matrix-form1}
\end{equation} 
where
\begin{equation}
\Omega_{\left(r,j\right)} 
=\omega_{n}^{2} -\frac{1}{r^{2} } 
\frac{d}{dr} \left(r^{2} \frac{d}{dr} \right)
+\frac{\left(j+1/2\right)^{2} }{r^{2} } 
+m^{2} 
\, 
\label{eq:Omega_diagonal-part}
\end{equation}
and
 $W(r) = (j +1/2)/r^2$.
Equation~(\ref{eq:Omega-Lambda_matrix-form1})
has a block form 
\begin{equation} 
\Lambda_{4}^{\left(j,r\right)} 
=
\left[\begin{array}{cc} 
 {\Lambda_{2}^{\left(j,r\right)} } 
  &
  {0} 
 \\
 {0}
 &
 {\Lambda _{2}^{\left(j,r\right)} } 
 \end{array}\right]
 \; ,
 \label{eq:Omega-Lambda_matrix-form2}
\end{equation} 
with the $2 \times 2$ matrix
\begin{equation}
  {\Lambda_{2}^{\left(j,r\right)} } 
  =
   \Omega_{\left(r,j\right)}
   \,
    {\mathbb {1}}_{2} 
 +
 W(r) \,
  \sigma_{1}
  = 
  {\left(
  \begin{array}{cc}
   \Omega_{\left(r,j\right)}  
  \;  & \; 
  W (r)
   \\ 
   W (r) 
 \;   & \; 
 {\Omega _{\left(r,j\right)} }
  \end{array}
  \right)} 
\; ,
\end{equation} 
which can be diagonalized with the 
coordinate-independent
unitary transformation
\begin{eqnarray}
V^{\dagger}
\,
 {\Lambda_{2}^{\left(j,r\right)} } 
\,
V
& = &
\check{\Lambda}_{2}^{\left(j,r\right)} 
\\
& = &
 \left(\begin{array}{cc} 
 {\Omega _{-\left(r,j\right)} } & {0} 
 \\ 
 {0} & {\Omega_{+\left(r,j\right)} }
 \end{array}
 \right)
 \, , 
\end{eqnarray}
where
\begin{equation}
 {\Omega _{\pm \left(r,j\right)} } 
 =
 \Omega _{\left(r,j\right)} 
 \pm 
 \frac{ ( j+1/2) }{r^{2} }
\; ,
\end{equation}
and
\begin{equation}
 V
  = 
 \frac{1}{\sqrt{2}}
  \begin{pmatrix}
  1  & 1
 \\
 -1 & 1
\end{pmatrix}
\; .
\end{equation}
Therefore,
\begin{equation}
\begin{array}{rcl} 
\det \Lambda _{4}^{\left(j,r\right)}
& {=} & 
\det ^{2} 
\left(
\Lambda _{2}^{\left(j,r\right)}
 \right)
=
\det ^{2} 
\left(
\check{\Lambda} _{2}^{\left(j,r\right)}
 \right)
 = 
 \det ^{2} 
 \left(\begin{array}{cc} 
 {\Omega _{-\left(r,j\right)} } & {0} 
 \\ 
 {0} & {\Omega_{+\left(r,j\right)} }
 \end{array}\right)
 \,  .
 \end{array} 
 \label{eq:determinant-Lambda}
\end{equation}

Now we are ready to evaluate the partition functional integral~(\ref{eq:Z_spherical3}), including the 
angular momentum degeneracy.  
As $j$ takes only semi-integer values 
$(j = 1/2, 3/2, 5/2, \ldots)$,
 let $l=j+{1\mathord{\left/ {\vphantom {1 2}} \right. \kern-\nulldelimiterspace} 2} =1,2,3 \ldots$. 
 With this notation,
calling  $\Omega _{\pm } \equiv \Omega_{\left(l,l-1\right)} $, i.e., $\Omega _{+} \equiv \Omega_{l}$
and $\Omega _{-} \equiv \Omega_{l-1}$, we have
\begin{equation} 
\Omega _{\pm } \equiv \Omega_{\left(l,l-1\right)} 
=\omega _{n}^{2} -\frac{1}{r^{2} } 
\frac{d}{dr} \left(r^{2} \frac{d}{dr} \right)+\frac{l\left(l\pm 1\right)}{r^{2} } + {m}^{2} . 
\end{equation} 
We can then write Eq.~(\ref{eq:Z_spherical3}) 
as 
\begin{equation} 
\displaystyle\prod _{n,l}\det \, \!^{2l}  \Lambda _{2}^{\left(l,r\right)}
 =\displaystyle\prod _{n,l=1}\left(\det \Omega _{l} \right)^{2l} 
  \displaystyle\prod _{n,l=1}\left(\det \Omega _{l-1} \right)^{2l}  . 
 \label{eq:Lambda-l_to_Omega-l} 
 \end{equation} 
In addition,
shifting the index and singling out one power of $\Omega_{l}$,
the following two relations ensue:
\begin{equation}
\displaystyle\prod _{l=1}\left(\det \Omega _{l-1} \right)^{2l} 
 =\displaystyle\prod _{l=0}\left(\det \Omega _{l} \right)^{2l+2} 
  =\displaystyle\prod _{l=0}\left(\det \Omega _{l} \right)^{2l+1} 
   \displaystyle\prod _{l=0}\det  \Omega _{l} \,, 
 \label{eq:Omega-l_aux1} 
 \end{equation} 
\begin{equation} 
\displaystyle\prod _{l=1}\left(\det \Omega _{l} \right)^{2l}  
=\displaystyle\prod _{l=0}\left(\det \Omega _{l} \right)^{2l}  . 
 \label{eq:Omega-l_aux2} 
 \end{equation} 
Therefore, from Eqs.~(\ref{eq:Lambda-l_to_Omega-l}) through (\ref{eq:Omega-l_aux2}),
the final expression for the spherical-coordinate representation of the 
partition function is obtained, 
\begin{equation} 
Z=\left(\displaystyle\prod _{n,l=0}\left(\det \Omega _{l} \right)^{2l+1}  \right)^{2} . 
\label{eq:Z_ang-momentum}
\end{equation} 
Now, from the standard resolution of the Laplacian operator in spherical coordinates,
$\displaystyle\prod _{n,l=0}\left(\det \Omega _{l} \right) ^{2l+1} $ 
is the spherical-coordinate factorization (up to a constant) of
 $\det \left(  \square _{E} + m^{2} \right)$,
 where 
$\square _{E} =- \left( \partial _{\tau }^{2} + \boldsymbol{\nabla }^{2} \right)$.
Therefore,
Eq.~\eqref{eq:Z_ang-momentum} 
then becomes, 
up to a constant, equal to the expression of Eq.~\eqref{eq:Z_momentum}, 
which was to be demonstrated.

\section{Conclusions}
 
We have shown,
via an explicit construction,
 the equivalence of the calculation 
for the partition function of a free gas of fermions 
in flat spacetime in Cartesian and in spherical coordinates. 
The latter involved an expansion of the fermion fields in ``generalized harmonics,'' as computed
in this paper. 
While the result is not surprising, our treatment of the path integral in spherical coordinates
highlights novel features and subtleties.
 It is technically remarkable for instance, to see the emergence of the zero mode ($l=0$)
  for the scalar case in the process of ``squaring'' the fermionic determinant. 
 
  Beyond its own merits, this flat-spacetime calculation lends strong support to the correctness 
  of the corresponding calculation in the case of a black hole background 
  performed recently by the authors.
  In that case,  Eq.~(\ref{eq:Z_spherical3}),
 after following the same procedure, is to be replaced by
a more complicated expression with additional terms and factors governed by the scale factor
$  f(r)
 = 1 -2M/r$ 
of the Schwarzschild metric
$  ds^{2} 
 =
 f\left(r\right)\left( dx^{0} \right)^{2}
 -\left[ f\left(r\right) \right]^{-1}
 dr^{2} 
 -r^{2} \left(d\theta ^{2} 
 +\sin ^{2} \theta \,  d\phi ^{2} \right)
 $
in $D=4$ spacetime dimensions
[where the
 Riemannian  spacetime geometry is described by a  metric with signature $\left(+,-,-,-\right)$].
As our computation shows, the determinant in Eq.~(\ref{eq:Z_spherical3})
has a  tight structure that allows for a full diagonalization into products of determinants 
over subspaces of the original one. 
 The NH expansion method is crucial to systematically
 isolate the leading contribution to the divergent part 
 of the free energy---hence of the entropy---of the fermionic thermal atmosphere 
 surrounding the black hole, which leads to the Bekenstein-Hawking law within the framework 
 of `t Hooft's brick-wall approach.
 Therefore,
  while the black-hole calculation is consistent
  with the expected fundamental result of black hole thermodynamics,
   it is important to establish
   the full validity of the technical aspects of the NH expansion, 
  and this paper lends further credibility to our approach.

\appendix

\section{Dirac Lagrangian and Dirac Matrices: 
\\
Miscellaneous Definitions and Conventions}
\label{sec-app:Euclidean_Dirac}

It is customary to define the Euclidean Lagrangian density 
${\mathcal L}_{E}$ from the Minkowskian one ${\mathcal L}$ via
\begin{equation}
{\mathcal L}_{E}
= -{\mathcal L} (t= -i \tau)
\; ,
 \label{eq:Euclidean-Lagrangian_recipe}
\end{equation}
so that the ordinary action $S=\int dt \int d^{3}x {\mathcal L}$ leads to
the Euclidean action $S_{E}  = -iS$.
This permits the transition from the ordinary time 
expression to the Euclidean one: $e^{iS} \longrightarrow e^{-S_{E}}$,
leading to a finite-temperature field theory for the partition
function with periodic or antiperiodic boundary conditions~\cite{TFT1a,TFT1b,TFT2}.

For a Dirac field with metric signature $(1,-1-1,-1)$, the Lagrangian reads
\begin{equation}
{\mathcal L} = 
\bar{\psi }
\left( i \gamma^{\mu } 
\partial _{\mu }
 - m\right)
 \psi 
 \, ,
 \label{eq:Dirac-Lagrangian_flat-Minkowski}
\end{equation}
so that the Euclidean Lagrangian and action of Eq.~(\ref{eq:Euclidean-action}) follow
with an appropriate redefinition of the Dirac matrices.

While the Euclidean definitions above for the action and Lagrangian are pretty much unique
(up to a choice of the metric signature), the conventions defining 
the
 Euclidean Dirac matrices are not.
 A convenient representation 
can be obtained by 
 ${\gamma_{E}^{0} =\gamma ^{0} } $ and
  ${\gamma_{E}^{j} =- i \gamma ^{j} }$ ($j=1,2,3$), i.e.,  
\begin{equation}
\left\{
\begin{array}{l} 
{\gamma_{E}^{0} =\gamma ^{0} } 
\\
 {\gamma_{E}^{1} =-i\gamma ^{1} } 
\\
 {\gamma_{E}^{2} =-i\gamma ^{2} } 
\\
 {\gamma_{E}^{3} =-i\gamma ^{3} ,} 
\end{array}
\right.
\label{eq:Euclidean_Dirac-matrices}
\end{equation} 
where,
 the $\gamma^{\mu }$'s 
are the usual Dirac matrices for Minkowski spacetime with
 signature $\left(+,-,-,-\right)$. 
 This choice guarantees
 that, with  the signature of Euclidean space
$\left(+,+,+,+\right)$,
the matrices satisfy the Euclidean version of the Clifford algebra, i.e., 
\begin{equation} 
\left\{\gamma_{E}^{\mu } ,\gamma_{E}^{\nu } \right\}=2\delta ^{\mu \nu }.  
\label{eq:Euclidean_Clifford-algebra}
\end{equation} 
Explicitly, in the Dirac representation we are using in this paper,  
\begin{eqnarray} 
\begin{array}{l} 
\gamma_{E}^{0} 
 = 
\left(\begin{array}{cccc} 
{1} & {0} & {0} & {0} 
\\ {0} & {1} & {0} & {0} 
\\ {0} & {0} & {-1} & {0}
\\ {0} & {0} & {0} & {-1} 
\end{array}\right)
\; \; \; \; \; \; \; \; \; \; \; \;
\gamma_{E}^{1}
 =
 \left(\begin{array}{cccc}
  {0} & {0} & {0} & {-i} 
 \\ {0} & {0} & {-i} & {0} 
 \\ {0} & {i} & {0} & {0}
  \\ {i} & {0} & {0} & {0} 
  \end{array}
  \right)
  \vspace*{0.15in}
  \\ 
  \gamma_{E}^{2}
  = 
 \left(\begin{array}{cccc} 
 {0} & {0} & {0} & {-1}
  \\ {0} & {0} & {1} & {0} 
  \\ {0} & {1} & {0} & {0}
  \\ {-1} & {0} & {0} & {0}
  \end{array}\right)
  \; \; \; \; \; \; \; \; \; \; \; \; 
  \gamma_{E}^{3} 
  =
  \left(\begin{array}{cccc} 
  {0} & {0} & {-i} & {0} 
  \\ {0} & {0} & {0} & {i} 
  \\ {i} & {0} & {0} & {0} 
  \\ {0} & {-i} & {0} & {0} 
  \end{array}\right)
  \, . 
  \end{array}
  \label{Euclidean-Dirac-matrices_Dirac-reps}
  \end{eqnarray} 
It should be noticed that other placements of the imaginary unit
in the components of Eq.~(\ref{eq:Euclidean_Dirac-matrices}) are possible to still satisfy
  Eq.~(\ref{eq:Euclidean_Clifford-algebra}); and that, alternatively,
   the opposite sign is often chosen
on the right-hand of   Eq.~(\ref{eq:Euclidean_Clifford-algebra}).

A few remarks on the covariant form of the Dirac equation are in order.
In any non-Cartesian coordinate frame, the Dirac operator takes 
the form~(\ref{eq:Dirac-Lagrangian_flat-Minkowski}) above, but with the replacement 
$\partial \rightarrow \slashed{\nabla}$, 
so that~\cite{Birrell-Davies_QFT-CST}
\begin{eqnarray}
{\mathcal L} 
& = & 
\bar{\psi }
\left( i
\slashed{\nabla} _{\mu }
 - m\right)
 \psi 
 \, ,
\\
\slashed{\nabla} 
& = & 
\gamma^{a} e^{\mu}_{\, a} \left( \partial_{\mu} - \frac{ i}{4} \omega_{\mu}^{\; \; ab}  \, \sigma_{ab} \right)
\; ,
\end{eqnarray}
where $\sigma_{ab} = i[\gamma_{a}, \gamma_{b}]/2$
 (rescaled commutator of the Dirac matrices,
related to spin)
and
$\omega_{\mu}^{\; \; ab}$ is the spin connection,\footnote{With the index positioning
convention
 $\omega_{\mu}^{\; \; \hat{\beta}\hat{\gamma}}  $, where $\mu$ is the coordinate index for covariant derivatives,
 as in Ref.~\citen{Carroll_GR}.
 One can then write $\omega_{\hat{\alpha}}^{\; \; \hat{\beta}\hat{\gamma}} 
 =
 e_{\; \; \hat{\alpha}}^{\mu}
 \omega_{\mu}^{\; \; \hat{\beta}\hat{\gamma}} 
 $, as displayed above in spherical coordinates,
 and also raise and lower indices with the Minkowskian metric as needed.}
with Latin indices  labeling the local frame and Greek indices labeling the coordinates.
The spin connection coefficients can be computed easily using either Cartan calculus or
the 
tetrad postulate (vanishing covariant derivative of $e_{\mu}^{\; \; a}$)
 and associated 
 derived algorithms~\cite{Carroll_GR}.
 For spherical polar coordinates, 
 using the orthonormal tetrad, with $a\equiv \hat{r}, \hat{\theta}, \hat{\phi}$,
  the only nonzero coefficients
  are 
 $\omega_{\hat{\theta}}^{\; \; \hat{\theta}\hat{r}} = - \omega_{\hat{\theta}}^{\; \; \hat{r} \hat{\theta} }
 =
 \omega_{\hat{\phi}}^{\; \; \hat{\phi}\hat{r}} = - \omega_{\hat{\phi}}^{\; \; \hat{r}\hat{\phi}} = -1/r
 $;
$ \omega_{\hat{\phi}}^{\; \; \hat{\phi}\hat{\theta}} = - \omega_{\hat{\phi}}^{\; \;  \hat{\theta}\hat{\phi}} = -(\cot \theta)/r $.
 As a result,
 after Euclideanization with the procedure described above, 
 the Dirac operator~(\ref{eq:Dirac-operator_angular-mod}) can be confirmed independently.

\section{Separation of Variables in Spherical Coordinates}
\label{sec-app:spherical_reduction}

Due to the structure of the Dirac operator, 
the separation into radial and angular components is more complicated 
than for the one-component scalar case. 
A convenient strategy
is as follows: Eq.~\eqref{eq:eigenvalue-eq_sphericalH} 
shows that a possible reduction 
 to a two-dimensional problem
 ensues by setting up
 the generalized eigenvalue 
problem\footnote{Notice that  $\lambda _{+}$ cannot be equal to $\lambda _{-}$.}
\begin{equation} \label{A.2.1)} 
\left(\left[\begin{array}{cc} 
{0} & {1} 
\\ 
{1} & {0} 
\end{array}
\right]\left(\partial _{\theta } 
+\frac{1}{2} \cot \theta \right)
+\left[\begin{array}{cc} {0} 
& {-i}
 \\ 
 {i} & {0} 
 \end{array}\right]
 \frac{1}{\sin \theta } \partial _{\phi } 
 \right)
 \left(\begin{array}{c}
  {Y_{1} }
   \\ {Y_{2} }
    \end{array}\right)
    =\left(\begin{array}{c} 
    {\lambda _{+} Y_{1} } 
    \\ {\lambda _{-} Y_{2} } 
    \end{array}
    \right) 
    \, ,
\end{equation} 
which involves two coupled differential equations.
These can easily be uncoupled by
a straightforward elimination of  one of the functions; for example,
solving first for $Y_{2}$ to obtain an equation for $Y_{1}$:
\begin{equation}
\label{3_14_} 
\biggl[
 \bigl(
 \partial _{\theta } 
+\frac{1}{2} \cot \theta 
\bigr)
-\frac{i}{\sin \theta } \partial _{\phi }
 \biggr]
\biggl[
\bigl(
\partial _{\theta } 
+\frac{1}{2} \cot \theta 
\bigr)
+\frac{i}{\sin \theta } \partial _{\phi } 
\biggr]
Y_{1}  
=
\lambda _{+} \lambda _{-} Y_{1} 
 \, .
\end{equation}
This equation can be solved by separation of variables
with building blocks
\begin{equation} \label{3_15_} 
Y_{1} (\theta ,\phi )=e^{im\phi } f(\theta )
\, , 
\end{equation}
where $f(\theta)$ satisfies the following equation:
\begin{equation} 
\frac{d^{2} f}{d\theta ^{2} } 
+\cot \theta \frac{df}{d\theta }
 -\biggl(
 \frac{1}{4} 
+\frac{1}{4\mathop{\sin }\nolimits^{2} \theta }
 +\frac{m^{2} }{\mathop{\sin }\nolimits^{2} \theta }
 -\frac{m\cos \theta }{\mathop{\sin }\nolimits^{2} \theta } 
 \biggr)
 f
 =\lambda _{+} \lambda _{-} f
 \, .
 \label{eq:Jacobi-DE_theta} 
\end{equation}

With the further change of variables,
\begin{equation}
\left\{
\begin{array}{l} 
{x=\cos \theta }
 \\
  {f (x)
  =
  (1-x)^{ (m-1/2)/2 } 
  \, 
  (1+x)^{ (m+1/2)/2 } 
    \,
  P(x)}
\, .
 \end{array}
\right.
\end{equation}
 one obtains the following equation for $P(x)$:
\begin{equation} 
(1-x^{2} )\frac{d^{2} P}{dx^{2} } 
+
\biggl[
1-(2m+2)x 
\biggr]
\frac{dP}{dx} 
+
\biggl[
-\frac{1}{4} -\lambda _{+} 
\lambda _{-} -m(m+1)
\biggr]
P=0
\, .
 \label{eq:Jacobi-DE} 
\end{equation}
This 
is the Jacobi differential equation~\cite{abr:72}; for the usual ``boundary conditions'' on
the angular variables, the acceptable solutions are  
 the Jacobi polynomials
$ P_{n}^{^{(\alpha ,\beta )} } (x)$,
with $n$ being a non-negative integer.
In terms of the parameters in Eq.~(\ref{eq:Jacobi-DE}),
 it follows that $\alpha, \beta= m \mp 1/2$;
 moreover
the requirement that $n$ be an integer
implies that the product
\begin{equation}
\lambda _{+} \lambda _{-} 
=-\left(j+1/2\right)^{2}
\end{equation}
 has to be a negative integer:
$\lambda _{+} \lambda _{-}  = -1, -2, -3,\ldots$  
(and hence $j$ a semi-integer:
$j = 1/2, 3/2, 5/2, \ldots$). 
We can then choose $\lambda _{\pm } =\pm \left(j+1/2\right)$. 
 Solving for $Y_{2}$ would give a similar equation to
 Eq.~(\ref{eq:Jacobi-DE})
  but 
  with the substitution
    \begin{equation}
    1-(2m+2)x 
    \longrightarrow
    - 1- (2m+ 2 ) x 
    \end{equation}
    in
    the coefficient of the first derivative.
        Therefore, we get
\begin{equation} \label{_3_19_} 
\left(\begin{array}{c}
 {Y_{1jm} } \\ {Y_{2jm} } 
 \end{array}\right)
 =
 C_{lm} e^{im\phi } 
 \left(\begin{array}{c} 
 {
  (1-x)^{ (m-1/2)/2 } 
  \, 
  (1+x)^{ (m+1/2)/2 } 
  \,
 P_{j-m}^{^{(m-1/2 ,m+1/2 )} } (x) 
 }
 \\
  {
 (1-x)^{ (m+1/2)/2 } 
  \, 
  (1+x)^{ (m-1/2)/2 } 
\,
 P_{j-m}^{^{(m+1/2 ,m-1/2 )} } (x)
 } 
 \end{array}\right) 
 \, .
\end{equation}
In addition, these eigenfunctions $Y$'s are chosen to be orthonormal,
\begin{equation} \label{A3)} 
\begin{array}{l}
 \iint  Y_{1jm_{j} }^{*} Y_{1j'm'_{j} } 
 \sin \theta d\theta d\phi 
=
\delta _{jj'} \delta _{m_{j} m'_{j} } 
\, 
\\
 \iint  Y_{2jm_{j} }^{*} Y_{2j'm'_{j} } 
\sin \theta d\theta d\phi =\delta _{jj'} \delta _{m_{j} m'_{j} } 
\,.
\end{array} 
\end{equation}
It should be noticed that the spinors above correspond to the 
spherical-coordinate vierbein; if rotated by a similarity transformation into the Cartesian vierbein,
they turn into the familiar spin spherical harmonics~\cite{Landau-Lifshitz_QED}
that consist of building blocks $Y_{l, m\pm 1/2}$, as shown in 
Refs.~\citen{Shishkin-Villalba_similarity1} and \citen{Shishkin-Villalba_similarity2},
where a similar method is used.

\end{document}